\documentclass{aa}
\usepackage{graphicx}
\usepackage{amsmath}
\begin{document}

\title{Fireballs from Quark Stars in the CFL Phase}
\subtitle{Application to Gamma Ray Bursters}

\author{R. Ouyed\inst{1}
\and R. Rapp\inst{2}
\and C. Vogt\inst{3}}

\institute{Department of Physics and Astronomy, University of Calgary,
Calgary, Canada
\and
Cyclotron Institute and Physics Department,
Texas A\&M University, College Station, TX 77843-3366, USA
\and
Nordic Institute for Theoretical Physics, Blegdamsvej 17,
DK-2100 Copenhagen, Denmark
}
\offprints{ouyed@phas.ucalgary.ca}

\date{Received/Accepted}

\abstract{
Recent studies of photon-generation mechanisms in the 
color-superconducting Color-Flavor Locked (CFL) phase of dense quark 
matter have found  $\gamma$-ray emissivities in excess of 
$\sim 10^{50}$~erg~cm$^{-3}$~s$^{-1}$ for temperatures in the 
10-30~MeV range. We suggest that this property can trigger 
$\gamma$-ray bursts (GRBs) and associated fireballs at the surface 
of hypothetical hot (newly born) quark 
stars with an energy release of up to $10^{48}$$-$$10^{50}$~erg 
within a fraction of a millisecond. 
If surrounded by an accretion disk following its formation, the star's
bursting activity can last from tens of milliseconds to hundreds of 
seconds releasing up to $10^{52}$~erg in total energy.
We discuss typical features of observed GRBs within our model and 
explain how quark stars in the CFL phase might constitute natural 
candidates for corresponding inner engines.  
}

\maketitle

\section{Introduction}
Nuclear matter at high density and small temperature ($T$) is expected 
to exhibit color-superconductivity (CSC), induced by quark 
pairing and condensation at the Fermi surface, with energy gaps 
$\Delta$$\simeq$100MeV (Rapp, Sch\"afer, Shuryak, \& Velkovsky 1998;
Alford, Rajagopal, \& Wilczek 1998) and associated critical 
temperatures $T_{\rm c}$$\simeq$0.6$\Delta$, above which 
thermal fluctuations destroy the condensate
(for a review, see, e.g., Rajagopal \& Wilczek 2000).
The experimental relevance of such matter mostly pertains to 
astrophysical objects, in particular compact stars. 
Besides its impact on the equation of state (Lugones \& Horvath 2002, 
Alford \& Reddy 2003), CSC could affect the emission 
spectrum of compact stars as encoded in its electro\-weak 
properties (Jaikumar, Prakash, \& Sch\"afer 2002; 
Reddy, Sadzikowski, \& Tachibana 2003; Vogt, Rapp, \& Ouyed 2004).  
If CSC extends to the surface of a hypothetical quark star, pulsed
photon emission has been suggested as a mechanism for a fireball in 
the $\gamma$-ray burst (GRB) context for the 2-flavour 
superconductor (2SC) in (Ouyed \& Sannino, 2002)\footnote{In the 
2SC phase, up and down quarks pair into a color-antitriplet, 
leaving the quarks of the remaining color unpaired, and with   
five of the eight gluons acquiring a mass. The 3 massless gluons 
possibly bind into light glueballs subject to fast decays into 
photons (Ouyed \& Sannino 2001).}. 
Whereas at moderate densities (and $T$=0) the existence of 
the 2SC is still under debate (Alford \& Rajagopal, 2002; 
Buballa et al. 2004) -- it could be superseeded by, e.g., 
crystalline phases (Alford, Bowers, \& Rajagopal, 2001; 
Rapp, Shuryak, \& Zahed, 2001) --, the Color-Flavor Locked (CFL) 
phase is the favored ground state at sufficiently high 
density (Alford, Rajagopal \& Wilczek, 1999). 
In CFL matter all three quark flavours (up, down and strange) have a mass 
which is negligible compared to the (quark) chemical potential, $\mu_q$,
so that they participate equally in the color condensation, 
breaking the full (local) color-symmetry and one (global) $SU(3)$ 
chiral symmetry.  Therefore, the zero-temperature CFL phase is   
electrically neutral without electrons, colorless, and its 
low-lying excitations are
characterised by  8+1 Goldstone bosons (due to chiral and baryon-number
symmetry breaking). 

In previous work (Vogt, Rapp, \& Ouyed, 2004), hereafter VRO, 
we have explored photon emission and absorption mechanisms in the 
CFL phase. Based on the Goldstone boson 
excitations (``generalised pions") of the broken chiral symmetry we 
have employed a Hidden Local Symmetry formalism including vector mesons 
(``generalised $\rho$-mesons") to assess photon production rates and 
mean free paths at temperatures suitable for CFL. We have 
found that above $T\simeq$~5-10~MeV, the emissivities
from pion annihilation, $\pi^+\pi^-\rightarrow \gamma\gamma$ and 
$\pi^+\pi^-\rightarrow \gamma$ (rendered possible due to an in-medium 
pion dispersion relation), dominate over those from conventional 
electromagnetic annihilation, $e^+e^-\to \gamma\gamma$. 
Given the very small photon mean free path at the temperatures
 of interest, $T$=10-30 MeV (see Fig. 3 in VRO), emission and absorption
 are in equilibrium and reflect the surface temperature of the star with
 a flux corresponding to that of a blackbody emitter.
 Therefore the pertinent cooling will largely depend on how efficiently
 heat conducted by the Goldstone-bosons from the interior can be emitted
 from the surface in form of the thermalized photons.
The objective of the present article is to 
(i) investigate the astrophysical consequences of such photon production
mechanisms\footnote{We note that our study is different from those
involving the $e^+e^-$ emission above the surface of CFL stars 
(see, e.g., Page \& Usov 2002 and references therein).
The latter scenarios reside on the fact that the 
color-superconductive matter of the star carries a charge so that
an $e^-$ abundance can build up thereby generating a critical electric 
field that in turn produces and emits $e^+e^-$ pairs and photons. 
This is quite different from our model which is based on photon 
generation inside the star with negligible $(e^+e^-)$ emission.} 
for the early cooling history of hot CFL stars (Sect.~\ref{sec:cooling}) 
and (ii) evaluate whether the resulting fireballs can drive/power GRBs
(Sect.~\ref{sec:grbs}). 
Conclusions are given in Sect.~\ref{sec:concl}.

\section{Early Cooling of CFL stars}
\label{sec:cooling}
\subsection{Plasma photon attenuation}
The surface emissivity of photons with energies below
$\hbar\omega_{\rm p}$~$\simeq$23~MeV ($\omega_{\rm p}$: 
electromagnetic plasma frequency) is strongly suppressed 
(Alcock, Farhi, \& Olinto 1986; Chmaj, Haensel, \& Slomi\'nsli 1991; 
Usov 2001). 
As shown in VRO, average photon energies in CFL matter at 
temperature $T$ are $\sim$3$T$. Therefore, as soon
as the surface temperature of the star cools
below $T_{\rm a} = \hbar\omega_{\rm p}/3\simeq 7.7$~MeV, 
we consider the photon emissivity to be shut off (``attenuated"), 
i.e., photon emission only lasts as long as the star cools from 
its initial temperature $T_0$ to $T_{\rm a}$. With this in 
mind we proceed to study the thermal evolution of the star. 

\subsection{Heat transfer and thermal evolution}
We describe the thermal evolution by a diffusion equation,
\begin{equation}
c_v \, \frac{\partial T}{\partial t}
=\frac{1}{r^2} \, \frac{\partial}{\partial r} \, \bigg( r^2 \, \kappa \,
\frac{\partial T}{\partial r} \bigg) \,,
\label{eq:diffusion1}
\end{equation}
where $c_{\rm v}$ and $\kappa$ are the specific heat and thermal 
conductivity of the star matter, respectively. In the CFL phase, these 
are dominated by massless Goldstone bosons, as evaluated  
in Jaikumar et al. (2002), 
\begin{equation}
 c_{\rm v} = \frac{2 \sqrt{3} \, \pi^2}{5} \, T^3
 = 7.8 \times 10^{16} \, T_{\rm MeV}^3\  {\rm erg\ cm^{-3}\ K^{-1}} \ ,
\label{eq:cv}
\end{equation}
and in Shovkovy \& Ellis (2002), 
\begin{equation}
 \kappa=1.2 \times 10^{27} \, T^3_{\rm MeV} \, \lambda_{GB} \,
 {\rm erg\ cm^{-1}\ s^{-1}\ K^{-1}} \,  
\label{eq:kappa}
\end{equation}
with a Goldstone-boson mean free path (in [cm] in Eq.~(\ref{eq:kappa})) 
\begin{equation}
 \lambda_{GB}(T) = \frac{4 (21-8 \ln 2)}{15 \sqrt{2} \, \pi \, T_{\rm MeV}}
 \, \exp\Bigg( \sqrt{\frac32} \frac{\Delta_{\rm MeV}}{T_{\rm MeV}} \Bigg) 
\ {\rm cm}\,  .
\label{eq:lambda}
\end{equation}
Effects of a crust have been ignored in the present study of the 
star's cooling history (see discussion in \S~\ref{sec:crust}).

In dimensionless units, the diffusion equation becomes
\begin{equation}
\frac{\partial \tilde{T}^4}{\partial \tau} = \frac{\alpha}{\tilde{r}^2}
\frac{\partial}{\partial \tilde{r}}\left( \tilde{\lambda}_{GB} 
\tilde{r}^2 \frac{\partial \tilde{T}^4}{\partial\tilde{r}}\right)\ 
\label{eq:diffusion2}
\end{equation}
with $\tilde{r}=r/R$, $\tilde{T}=T/T_0$, $\tau = t/(R/c)$,
$\tilde{\lambda}_{GB}=\lambda_{GB}/R$ and a numerical 
coefficient $\alpha$=0.512 ($c$ is the speed of light).
As initial condition we assume the temperature of the star at 
$\tau=0$ be uniform, $\tilde{T}(\tau=0)=\tilde{T}_0$. 
We further have to impose boundary conditions for the temperature 
gradient in terms of the heat flux,
$F(\tilde{r}) = -\tilde{\lambda}_{GB} \, 
\partial \tilde{T} / \partial \tilde{r}$, at the
center and the surface of the star,
 $F(\tilde{r}=0) = 0$ and $F(\tilde{r}=1) = \tilde{T}^4$.
Due to the large opacity we expect
that the photons to be thermalised upon leaving the star. Thus, 
we assume black-body radiation to be a good approximation (supported
by VRO).  The energy flux per unit time through a spherical shell
at a radius $\tilde{r}$ is evolved by taking into
account the gradient of temperature on both sides of the shell. 
For the outermost shell, the flux exits the star
via thermal photons, implying that 
the cooling curves result from photon emission only. 

\begin{figure}[!t]
\includegraphics[width=0.5\textwidth, angle=0]{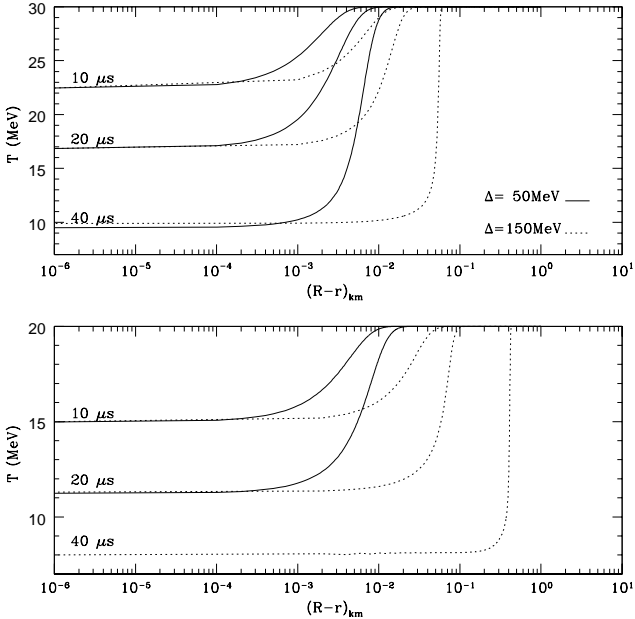}
\caption{Radial temperature profiles of CFL stars of radius $R$ 
according to solutions of the diffusion Eq.~(\ref{eq:diffusion1}) 
($r$: distance to the star's center) for two 
different gaps, $\Delta$=50~MeV (solid lines)
150~MeV (dotted lines), and initial temperatures
$T_0$=30~MeV (upper panel) and 20~MeV (lower panel).
In the lower panel, the $\Delta$=50~MeV case cools to 
$T_{\rm a}$=7.7~MeV (lower limit of the plots)
in less than $40~\mu$s.}
\label{fig:tempvsradius}
\end{figure}

\begin{figure}[!t]
\includegraphics[width=0.5\textwidth, angle=0]{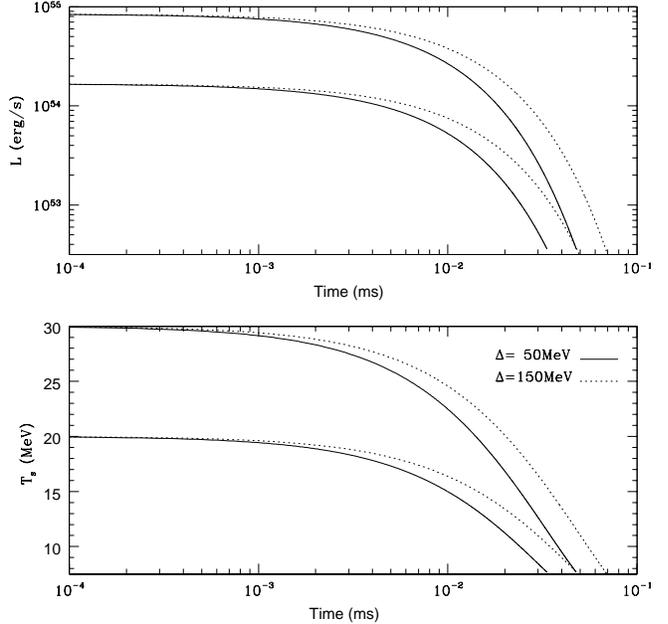}
\caption{Time dependence of surface temperature (lower panel) and
photon luminosity (upper panel) of a CFL star for two initial
temperatures, $T_0$=20 and 30~MeV (lower and upper lines,
respectively). Solid (dotted) lines are for $\Delta$=50(150)~MeV.}
\label{fig:tempvstime}
\end{figure}

Numerical solutions to Eq.~(\ref{eq:diffusion2}), using a 
finite-difference method, are summarized in 
Figs.~\ref{fig:tempvsradius} and \ref{fig:tempvstime} for a quark star 
of radius $R$=10~km. Fig.~\ref{fig:tempvsradius} shows the temperature 
vs. distance from the star's surface, $R-r$, indicating that $T$ drops 
to $T_{\rm a}$ in less than $0.1$~ms over a shell of thickness
less than $\sim$1~km. Neglecting heat transport,  
the cooling timescale may be roughly estimated from 
$c_v(\partial T /\partial t)$=$-\epsilon$=$-L/\Delta V$ 
($\Delta V$=$4\pi R^2 \Delta R$:  volume of the cooling 
shell of thickness $\Delta R$). Assuming a blackbody luminosity,
$L=4\pi R^2\sigma T^4$, one finds $t_{\rm cool}\simeq 
0.1{\rm ms} \times \Delta R_{\rm km} \times (\Delta T/T)$,
($\Delta T=T_0-T_{\rm a}$), approximately reflecting 
the numerical results.
Note that our estimate improves toward small $T$ and large
$\Delta$ for which $\lambda_{GB}$ is larger, so that
neglecting the delay time due to heat transport 
(Eq.~(\ref{eq:diffusion1})) is better justified;   
e.g., for $T_0$=20~MeV and $\Delta$=150~MeV (in which case
$\lambda_{GB}$=10$^{-2}$km, reaching 1~km at $T$$\simeq$15~MeV), 
$\Delta R_{\rm km}$$\simeq$1 (lower panel in 
Fig.~\ref{fig:tempvsradius}); with $(\Delta T/T)$$\simeq$1
one recovers $t_{cool}$$\simeq$0.1~ms, consistent with   
the lower panel of Fig.~\ref{fig:tempvstime}.

For identical $T_0$, a larger gap implies 
cooling deeper into the star, but slower in terms of the reduction
in surface temperature, cf.~Fig.~\ref{fig:tempvstime}.  
Again, this follows from the larger $\lambda_{GB}$, 
for which heat emerges from deeper in the star
and thus provides a larger energy reservoir.

When increasing the initial temperature, $T_0$, from 20~MeV to 30~MeV, 
the surface cooling does not change radically.
The main difference is that the cooling of the bulk sets in earlier 
so that the temperature gradient between surface and bulk is washed out 
in shorter time.

We recall that contributions from neutrino ($\nu$) emission are
not included in our analysis. By assuming
 uniform star density and temperature as our initial state
the consequences as shown in VRO are (i) $\nu$ emission from the surface
 is usually negligible in comparison
to photon emission in the temperature region we are considering (see
 \S 3.4 in VRO for more details), and (ii) $\nu$ emission from the bulk
will set in once the neutrino mean-free-path becomes comparable
to the star radius which is expected to occur for temperatures
below $\sim$5~MeV (Reddy et al. 2002); above, the neutrinos are
essentially trapped inside the star. This point is further discussed
in the conclusion.

We also note that photons will be redshifted
as they stream outwards due to the star's gravitational potential. 
First order General Relativity effects on the emissivities 
can be introduced through redshift factors 
expressed in terms of the star mass and radius.
We estimate the pertinent redshift reduction to be of the order of 
10\% (cf.~Frolov \& Lee 2005 for a recent analysis).
While the gravitational redshift will degrade the total energy 
emitted it should not affect the qualitative features of our
results.

\section{Application to Gamma Ray Bursts}
\label{sec:grbs}
The total energy released during 
cooling is computed from Fig.~\ref{fig:tempvstime}
as $E_{\rm tot}=\int L_{\gamma} dt$.
Thus, a CFL star with initial temperature
of 10-30~MeV can release on average $10^{48}$-$10^{50}$~erg 
within $\sim$0.1~ms as it cools to $T_{\rm a}$. 
We have at hand an engine that could be driving 
observed GRBs.

\subsection{Accretion disk and temporal variability}
\label{sec:variability}
Our main idea for the following is that the input energy to the
engine is provided by infalling matter from an accretion disk. 
The energy released by the hot CFL star translates into
$(0.01-1)M_{\odot}c^2$~s$^{-1}$ in accretion energy.
These values are reminiscent of hyperaccreting disks
(Popham, Woosley, \& Fryer 1999) suggesting that if the latter is 
indeed formed around a CFL star it could reheat the surface
of the star via accretion keeping the engine active for a much 
longer time and providing higher energies as compared to
collapse events only. More precisely, the total available energy 
is related to the disk mass, $M_{\rm disk}$, by
\begin{equation}
E_{\rm T} \simeq \eta M_{\rm disk} c^2
+\frac{M_{\rm disk}}{m_{\rm H}}\Delta E_{\rm B} \ ,   
\label{eq:etotal}
\end{equation}
where the first term accounts for the release in gravitational
binding energy of the accreted matter, with $\eta$$\sim$0.1 
(e.g., Frank, King, \& Raine 1992). 
The second term represents the binding energy released 
in the conversion from nuclear to CFL matter ($\Delta E_{\rm B}$ per 
nucleon). It is presumably emitted
via neutrinos (Glendenning 1997) so 
that $E_{\rm T}$$\simeq$$\eta M_{\rm disk}c^2$ remains 
available to photon production. 

\begin{figure*}[t!]
\includegraphics[width=0.9\textwidth, height=0.45\textwidth]
{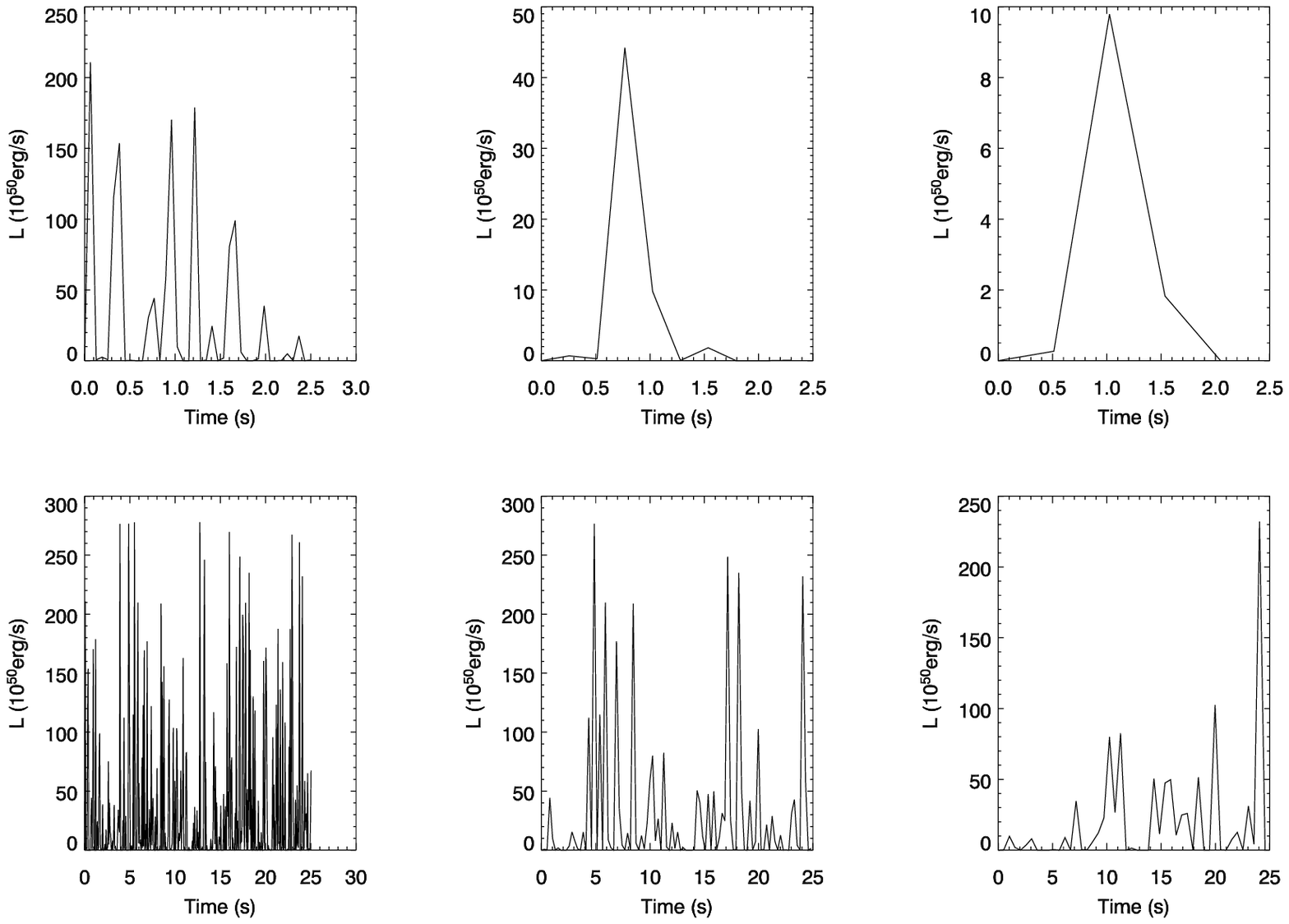}
\caption{Model results for the effects of a hyperaccreting disk on the
star luminosity; shown is the $\gamma$-ray burst emission versus time 
for three different time resolutions (64~ms, 256~ms, and 512~ms from 
left to right) representative of GRBs detectors 
(www.batse.msfc.nasa.gov/batse/grb/lightcurve/).
The upper (lower) panels are for a total accreted energy  
$E_T$=0.001$M_{\odot}c^2$ ($E_T$=0.01$M_{\odot}c^2$) 
with $T_{\rm max}$=30~MeV. 
}
\label{fig:variability30}
\includegraphics[width=0.9\textwidth, height=0.45\textwidth]
{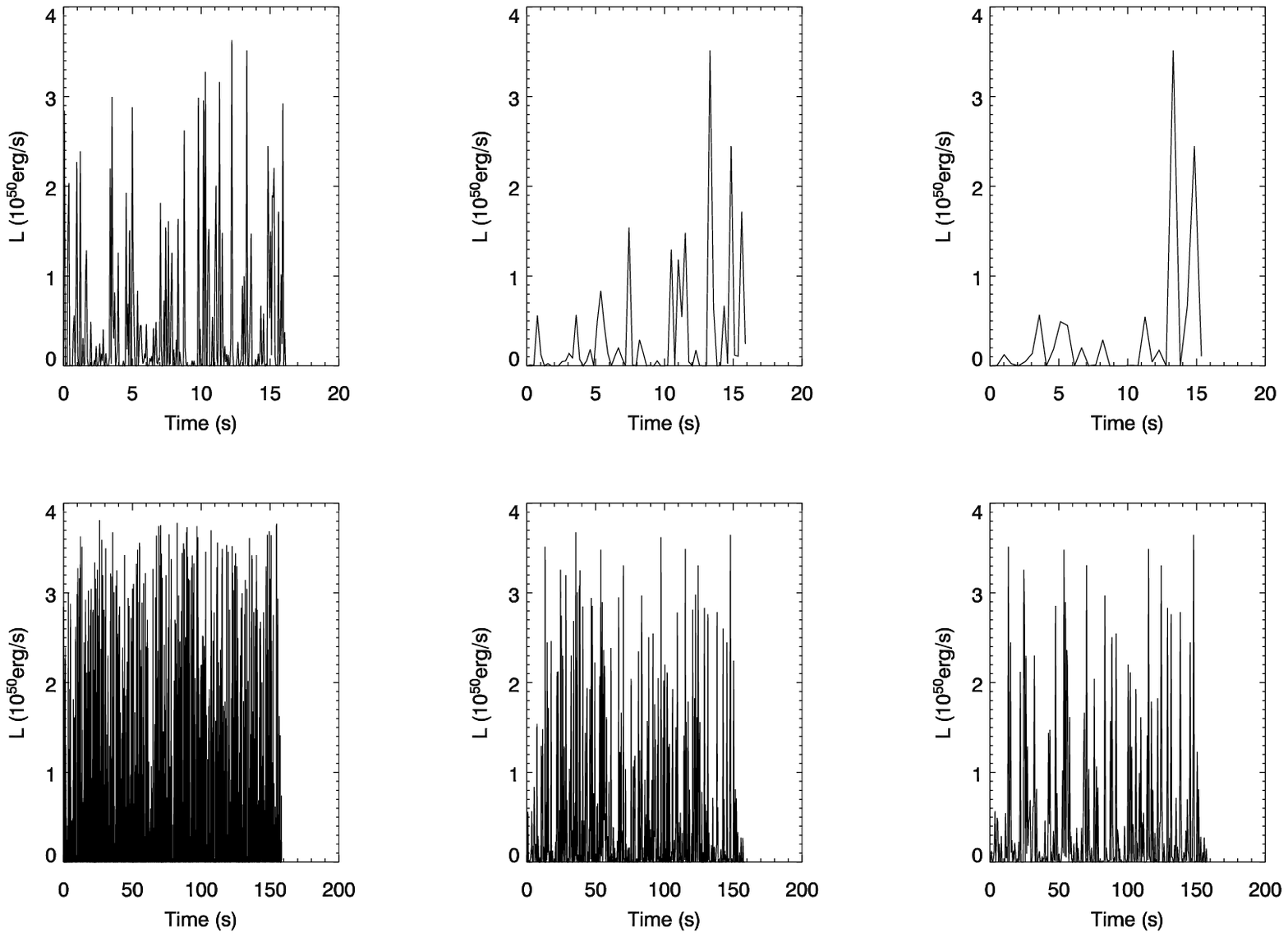}
\caption{Same as in Fig.~\ref{fig:variability30} but for 
$T_{\rm max}$=20~MeV. More sample of lightcurves can be found 
at {\sl http://www.capca.ucalgary.ca/}.
}
\label{fig:variability20}
\end{figure*}

We are modelling the effects of the accreted material on the 
CFL star in 
a very simplistic approach, as a random increase of the CFL surface 
temperature $T_s$ (heating) to a value $T_{\rm peak}$
with a uniform distribution
between the lower limit, $T_a$$\simeq$7.7~MeV, and an upper limit, 
$T_{max}$=15-30MeV. 
This roughly corresponds to expected average disk temperatures up to 
$T_{disk}$$\simeq$15~MeV or so (Popham et al. 1999); 
we neglect cooling processes during accretion.
Note that for accreted material with temperature 
below $T_{\rm a}$ there is no bursting and the engine remains shut 
off until further accretion drives $T_s$ above $T_{\rm a}$.
In these cases the accretion proceeds for longer than the average
accretion timescale.

\begin{figure*}[t!]
\includegraphics[width=0.9\textwidth, height=0.5\textwidth]
{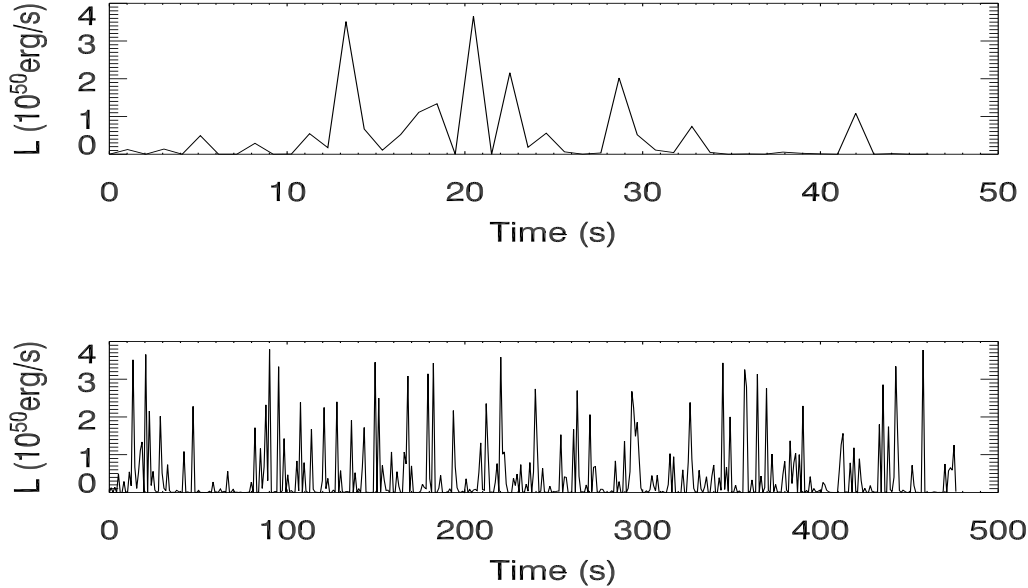}
\caption{$\gamma$-ray burst emission spectra versus time for a 
CFL star with accretion disk of mass $0.01M_{\odot}$ (or $E_{\rm T}=0.001M_{\odot}c^2$; upper panel) and $0.1M_{\odot}$ (or $E_{\rm T}=0.01M_{\odot}c^2$; lower panel), 
with $T_{\rm max}$=15~MeV and time resolution $1024$~ms.
More examples can be found at {\sl http://www.capca.ucalgary.ca/}.
}
\label{fig:variability_m15}
\end{figure*}

For the time intervals of accretion, $\Delta t_{\rm accr}$, we 
assume the free-fall 
time for a hyperaccreting disk as the relevant scale, 
$t_{\rm ff}$=$1/\sqrt{4\pi G \rho_{\rm disk}}$$\simeq$1~ms,
with $\rho_{\rm disk}$$\simeq$$10^{12}$~g~cm$^{-3}$ the
accretion disk density and $G$ the gravitational constant.
The accretion time intervals are sampled stochastically between 1
and 10~ms. The time scale, $\Delta t_{\gamma}$, of the 
subsequent $\gamma$-ray burst (cooling) 
follows from the diffusion Eq.~(\ref{eq:diffusion2})
as the time it takes to cool the surface to $T_a$.
During this time the accretion energy is transformed to photon 
energy which in a simple form can be written as 
$L_{\gamma}\Delta t_{\gamma}$=$\eta \dot{M}_{\rm disk}c^2
\Delta t_{\rm accr}$. 
Thus, each event (reheating and cooling) lasts for 
$\Delta t_{\rm event}$=$\Delta t_{\rm accr}+\Delta t_{\gamma}$,
for given $\Delta$ and $T_{\rm peak}$ (we set $\Delta$=50~MeV).   
Consequently, each episode lasts for a few milliseconds 
consisting of a linear increase of $T_s$ to 
$T_{\rm peak}$ followed by a rather sharp decrease to $T_{\rm a}$. 

For a given disk mass (see \S~\ref{sec:formation}), 
the simulation is carried out 
until the total available energy, $E_{\rm T}$$\simeq$$0.1 M_{\rm disk}c^2$ 
(Eq.~(\ref{eq:etotal})), is consumed.  
The resulting variability (displayed in 
Figs.~\ref{fig:variability30} and \ref{fig:variability20}
for $E_{\rm T}$=0.001$M_{\odot}c^2$, 0.01$M_{\odot}c^2$
and typical time resolutions of GRB detectors) share 
interesting similarities to those observed in GRB data, 
see, e.g., http://www.batse.msfc.nasa.gov\-/batse/grb/lightcurve.
Note that our spectra correspond to the activity directly at the 
engine while the GRB curves represent emission beyond the 
transparency radius (see, e.g., Piran 2000). Nevertheless, the engine 
activity should reflect the variability at large distances.

\subsection{Total duration}
\label{sec:duration}
Our simulations of the accretion suggest engine activities varying from 
milliseconds to hundreds of seconds depending on $M_{\rm disk}$ and 
$T_{\rm max}$ (Figs.~\ref{fig:variability30} and 
\ref{fig:variability20}). 
The average duration can be estimated as
$t_{\rm dur}$$\simeq$$\Delta \bar{t}_{\rm accr}\times 
(E_{\rm T}/\bar{E}_{\gamma})$
with average burst energy and
accretion timescale 
$\bar{E}_{\gamma}$$\simeq$$10^{49}$~erg and 
$\Delta \bar{t}_{\rm accr}\simeq$5~ms, respectively. 
One obtains $t_{\rm dur}$$\simeq$10~s
for a CFL star surrounded by a $0.1M_{\odot}$ disk.
For this disk mass
(corresponding to $E_{\rm T}$=$0.01M_{\odot}c^2$) the engine
activity can last for hundreds of seconds if $T_{\rm max}$ is 
sufficiently low (e.g., 15~MeV as in Fig.~\ref{fig:variability_m15}). 
In general, for a given $M_{\rm disk}$, smaller 
$T_{\rm max}$ imply smaller average energies released per episode 
(i.e. smaller accretion rates), and thus longer lifetimes.

We also note that the variability observed in the first few seconds 
of the $E_{\rm T}$=0.01$M_{\odot}c^2$ simulations is very similar to the  
$E_{\rm T}$=0.001$M_{\odot}c^2$ case which lasts for $\sim$2~s, 
recall Figs.~\ref{fig:variability30} and \ref{fig:variability20}.
Similarly, the variability seen in the first few seconds of a 
$E_{\rm T}$=0.1$M_{\odot}c^2$ simulation (not shown here) bears many
similarities to the $E_T$=0.01$M_{\odot}c^2$ case. 
This ``self-similar" behavior is due to  
the duration being controlled by accretion timescales 
$\Delta t_{\rm accr}$$>$$\Delta t_{\gamma}$, in connection with the 
stochastic accretion process whose main effect is to spread the 
events (peaks) randomly in time.

\subsection{Total energy}
In our model up to 10\% of $M_{\rm disk}c^2$ can be transformed into 
fireball energy. E.g., if an accretion disk with mass $0.1M_{\odot}$ 
surrounds a CFL star following its formation (see 
\S~\ref{sec:formation}), up to $10^{52}$ ergs
of energy can be released in $\gamma$-rays. Larger
disks lead to higher energy release but in this
case the star is likely to become a black hole in the process.

\subsection{Baryon loading and beaming}
\label{sec:crust}
The intense, localised (source size $<$100~km) and brief explosion 
implied by the observed GRB fluxes
imply the formation of a $e^+e^-$-photon fireball.
Furthermore, most of the spectral energy in GRBs
is observed at $\ge$0.5~MeV, so the optical depth
for $\gamma\gamma\rightarrow e^+e^-$ processes is very large.
Any photon generated above $0.511$ MeV will be degraded to below
0.511~MeV via the $\gamma\gamma\rightarrow e^+e^-$ process.
This is the so-called compactness problem leading
to a thermalised fireball (Ruderman 1975) unlike the optically
thin spectra observed in GRBs. 

The compactness problem can be resolved if the
emitting matter is moving relativistically toward the observer.
In this case the relative angle at which the photons
collide must be less than the inverse of the bulk
Lorentz factor, $1/\Gamma$, to effectively reduce the threshold
energy for $e^+e^-$ pair production (Goodman 1986). 
One can show that $\Gamma$~$\ge$~100 is required to
overcome the compactness problem (Shemi \& Piran 1990; Paczy\'nski 1990).
This relativistic outflow presumably arises from an initial
energy $E_0$ imparted to a mass $M_0$~$\ll$~$E_0/c^2$ close
to the central engine (cf.~M\'esz\'aros 2002 for more details).

Assuming the star is bare\footnote{It has been suggested that the plausible surface depletion of s quarks
 induces a Coulomb barrier allowing the presence
of a crust (Usov 2004). Nevertheless a crust should
be blown away by the  extreme radiation pressure induced by the
extreme temperature (tens of MeV) of the star immediately
following its birth and during its early history (e.g. Woosley\&Baron 1992).
We thus assume that the star is bare for most of
the bursting era where the high temperatures are maintained
by the infalling disk material.}
in our model the emitted photons are likely to interact   
with particles from the accreting material. To enable 
acceleration to Lorentz factors above 100 the photon energy must 
be imparted on particles representing on average less than 0.1\% 
of the total matter accreted per episode, since 
$\Gamma$=$\eta\dot{m}_{\rm acc.}c^2 \Delta t_{\rm accr}/m_{\rm ejec}
c^2$=$\eta (m_{\rm accr}/m_{\rm ejec})$.
In other words, we require most of the infall material to 
convert to CFL matter with only a small portion of it to be ejected.
This seems not unreasonable as most of the accreted particles
will instantly deconfine to quark matter upon 
contact with the star's surface (see, e.g., Weber 2005).

The amount of mass ejected will
vary from one episode to another leading to a spread 
in the Lorentz-factor distribution as already
indicated in Figs.~\ref{fig:variability30}, \ref{fig:variability20}
and \ref{fig:variability_m15}.
A fast loaded fireball that was injected after a slower one will
eventually catch up and collide,  converting some of the kinetic energy
of the shells to thermal energy.
This is reminiscent of the popular internal-shock model for GRBs 
(see, e.g., Kobayashi et al. 1997; Piran 2005) where a succession 
of relativistic shells collide and release the shock energy  
via synchrotron emission and inverse Compton scattering.

\begin{figure}[t!]
\includegraphics[width=0.45\textwidth]{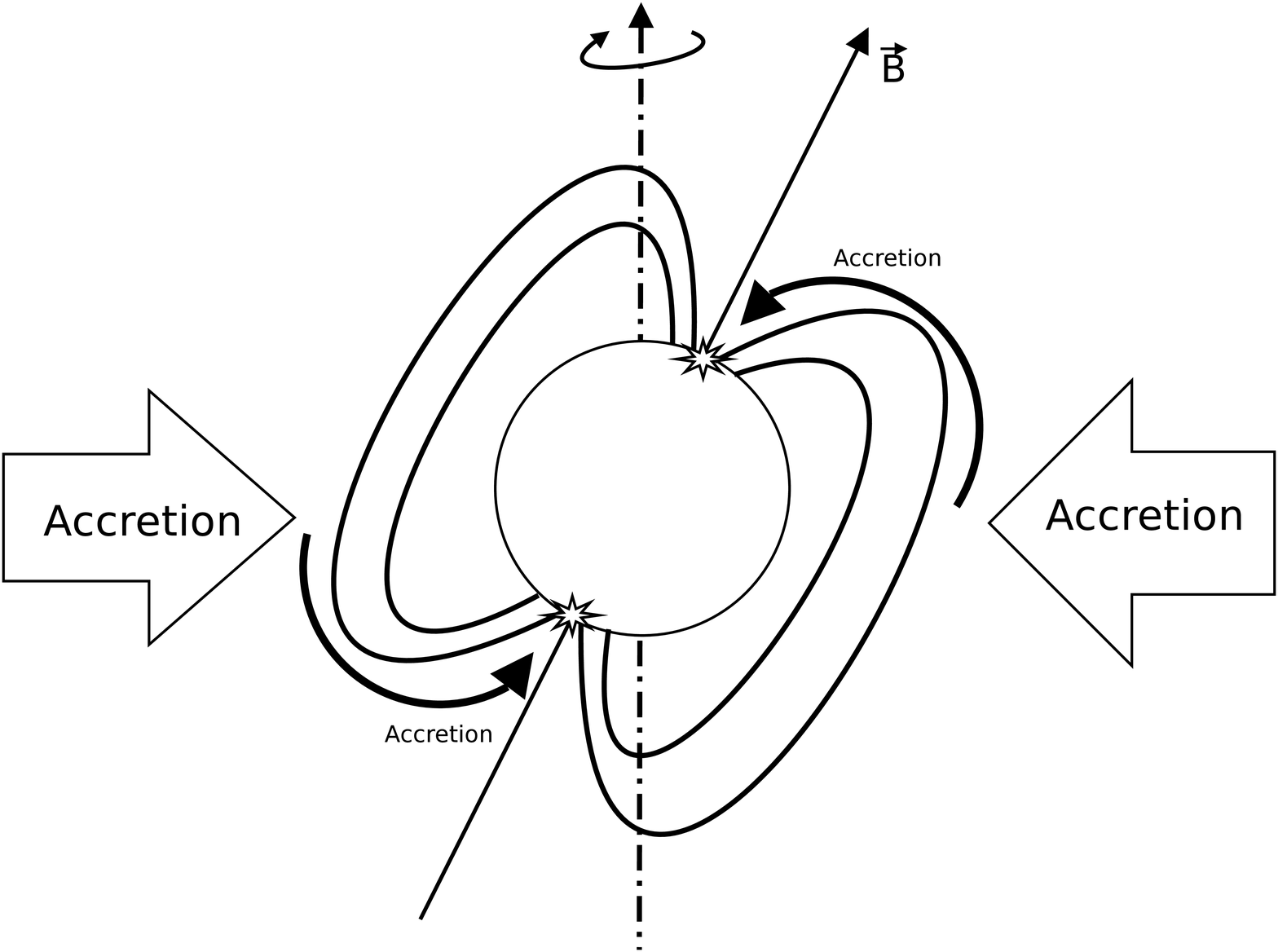}\\
\includegraphics[width=0.45\textwidth]{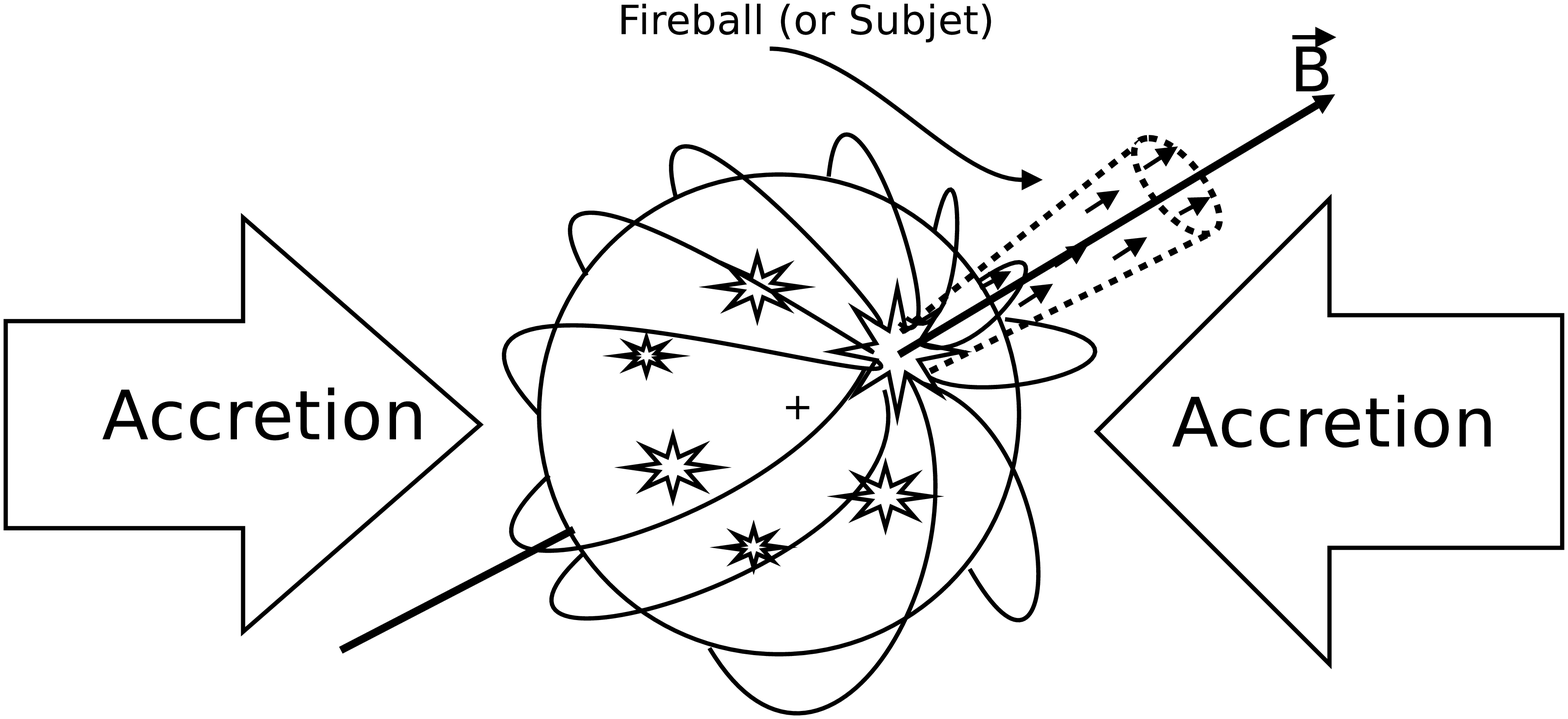}
\caption{Illustration of the intermittent accretion-ejection
mechanism in our model. The upper panel shows the disk
material channelled by the magnetic field toward the polar
regions reheating the surface temperature to $T_{\rm peak}$
(the flashy spots). When $T_{\rm peak} > T_{\rm a}$, 
a photon burst is triggered ejecting particles from the
accreting material. From a viewpoint along 
the polar axis (depicted by a cross in the lower panel),
the multiple flashes on the surface
of the star illustrate the locations of previous bursts 
(or ejections) induced by previous accretion episodes. 
The spatial spread of these flashes portrays the randomness in the
accretion and cooling timescales.
}
\label{fig:beaming}
\end{figure}

The intricate details of the accretion-ejection cycles 
are beyond the scope of this paper. Qualitatively, upon 
taking into account the star-disk magnetic field, 
we expect accretion to be channelled toward the polar regions as 
illustrated in Fig.~\ref{fig:beaming}. 
A refined scenario could further include hot bursting spots on 
the surface of the CFL star emitting $\gamma$-rays; the outgoing photons 
 interact with the particles from the accreting matter ejecting part of it. 
Thus, in our model, the loaded fireballs emanate mostly from the polar regions with 
collimation reinforced by the magnetic field (e.g., Fendt\&Ouyed 2004).
When a hot spot cools below $T_{\rm a}$, further accretion
is triggered leading to another photon-bursting spot on the
star as pictured in Fig.~\ref{fig:beaming}. The
time delay between two successive spots is related
to the randomness in the accretion (reheating)
and the cooling timescales as described in \S~\ref{sec:variability}.
This process continues until $E_{\rm T}$ is consumed, which
in some cases can support up to hundreds of episodes (subjets), 
as in Fig.~\ref{fig:variability_m15}.

\subsection{Formation scenarios/sites}
\label{sec:formation}
In scenarios for the formation of CFL stars massive progenitors 
are naturally favoured since they are more likely to lead to compact 
remnants with high enough densities in the core for a (phase) 
transition to quark matter to occur. This is in line with afterglow 
observations which have provided several hints that some GRBs are 
associated with massive star progenitors (Galama et al. 1998, 
van Paradijs et al. 1999). If the quark star forms
immediately following a supernova (SN) explosion of a massive 
progenitor, a thick disk ($M_{\rm disk}$~$>$~0.1$M_{\odot}$) is 
expected to form from fall-back material.
The faster fireballs from the CFL star 
will catch up with the preceding supernova shell
as to energise it. In fact, the energetic subjets from the
CFL star are capable of destroying the
original symmetry of the expanding SN shell.
Another possibility involving massive progenitors  
is that of a collapsar-like event (Woosley 1993)
where a quark star is formed instead of a black hole. In this
scenario as well one would favour thick and massive disks to form
around the star.

The hadron-quark transition may also happen long after 
the supernova explosion as in the Quark-Nova model
where a neutron star (originally formed from a massive progenitor)
can reach quark-matter densities through accretion from a companion,
or due to spin-down (Ouyed, Dey, \& Dey 2002; Ouyed
et al. 2004). In some cases, the phase transition may take
tens of millions of year to occur.
During the explosion up to $0.01M_{\odot}$
of material can be ejected (Ker\"anen, Ouyed, \& Jaikumar 2005)
which later turns into a hyperaccreting disk (Ker\"anen\&Ouyed 2003).

\subsection{Bimodality: Short and long duration GRBs}
The durations of GRBs observed by BATSE show
a bimodal distribution, which has led to a classification
of GRBs into short (with $t_{90}$~$<$~2~s, where $t_{90}$ is
the decay time to 10\% intensity) and long ($t_{90}$~$>$~2~s)
(Kouveliotou et al. 1993; McBreen et al. 1994).
It has been suggested that different underlying engines 
are operative, the short
ones being related to binary neutron star mergers, and
the long ones to the collapse of massive
stars (see M\'esz\'aros 2002, and references therein).
However, recent comparisons of short
GRBs lightcurves to the first few seconds
in long GRBs indicate that the two classes could be similar 
(e.g., Nakar\&Piran 2002; Ghirlanda et al. 2004).

As noted in \S~\ref{sec:variability},
our model provides a kind of ``self-similar" behaviour
where the variability in the first few seconds of a long duration
GRBs (thick disk) is reminiscent of that
of a short GRB (thin disk), both driven by the same engine.
As for the observed bimodality, it is possible that the formation 
scenarios discussed in \S~\ref{sec:formation} (namely, collapsar-like 
events and Quark-Nova explosion) lead to 2 distinct classes of 
accretion disk (thick with $M_{\rm disk}>0.01M_{\odot}$ and thin with 
$M_{\rm disk}< 0.01M_{\odot}$) surrounding the CFL star.

Interestingly, it has been argued recently in the literature that the
bimodality in GRB duration originates from discrete emission
regions (subjets) in the GRB jet (Toma, Yamazaki, \& Nakamura 2005
and references therein). In this model the multiplicity
of the subjets ($n_{s}$) along a line-of-sight differentiates between
short GRBs ($n_{s}\sim 1$) and long GRBs ($n_{s}\gg 1$). 
Many subjets have to be randomly launched
by the central engine for the model to work.
This bears a close resemblance to the picture we developed
here (cf.~Fig.~\ref{fig:variability_m15}) and warrants more
detailed studies of accretion-ejection within our model.

\section{Conclusion}
\label{sec:concl}
We have studied consequences of previously 
calculated photon ($\gamma$-ray) emission mechanisms in CFL matter
for the early cooling of CFL stars. Based on the notion that
pertinent emissivities essentially saturate the blackbody limit
for temperatures $T$~$\simeq$~10-30~MeV (due to underlying processes
involving the Goldstone bosons of CFL matter), 
we have solved a diffusion equation and found that each photon 
burst can release an average 
energy of $\sim$$10^{49}$~erg during a fraction of a millisecond.
We have suggested a schematic model within which the time 
variability and the long activity (up to hundreds of seconds) of 
a GRB engine is driven by a surrounding hyperaccreting disk resulting
from the formation process of the CFL star. This model reproduces
several features of observed GRB spectra.

In our simplified picture we assume that photons and
neutrinos are emitted at the same temperature (see
 section 3.4 in Vogt et al. for a discussion).
This certainly is far from a realistic situation where
 temperature gradients will be present and where
 the neutrinosphere will be buried deeper in the star at hotter
temperatures than the photosphere.  In this case neutrinos will dominate
 the cooling leading to timescale shorter than the 0.1 ms
 range we found when only photons are at play. This is of particular
 importance in the early stages immediately following the formation of the star.
  On the longer timescales, and within the  GRB context, the
 reheating mechanism (induced by the infalling accretion disk material)
 affects mostly the surface layers and thus specifically
 the photosphere region. A decoupling between the photosphere and
 the neutrinosphere is to be expected justifying the neglect of neutrinos
 cooling in the subsequent photon bursts/episodes.

For a more quantitative description of the complex energetics and 
dynamics involved in our model, advanced numerical simulations should 
be performed accounting for general relativistic effects, as well as 
the star-disk magnetic field. Such studies will be similar to what 
has been done in the case of black-hole-accretion-disk
systems (De Villiers, Staff, \& Ouyed 2005), with the black hole being
replaced by a CFL star.

\begin{acknowledgements}
The research of R.O. is supported by grants from the Natural Science 
and Engineering Research Council of Canada (NSERC) and the
Alberta Ingenuity Fund (AIF), and the research of R.R. is 
supported in part by a U.S. National Science Foundation CAREER
award under grant PHY-0449489.
\end{acknowledgements}

\end{document}